\UseRawInputEncoding
\documentclass[pra,aps,twocolumn]{revtex4}
\usepackage{amsmath,mathrsfs,amsbsy,color,bm,amsthm,amsfonts,epsfig,graphicx}
\usepackage{units}
\usepackage{bbm}
\usepackage{times}

\newcommand{\tr}{{\rm Tr}}

\newcommand{\ket}[1]{|#1\rangle}
\newcommand{\bra}[1]{\langle #1|}
\newcommand{\proj}[1]{\ket{#1}\bra{#1}}

\begin{document}

\title{Multipartite entanglement detection based on generalized state-dependent entropic uncertainty relation for multiple measurements}

\author{Li-Hang Ren$^1$}
\email{renlihang@hebtu.edu.cn}
\author{Yun-Hao Shi$^{2,3}$}
\author{Jin-Jun Chen$^4$}
\author{Heng Fan$^{2,3,5}$}
\email{hfan@iphy.ac.cn}

\affiliation{$^1$ College of Physics and Hebei Key Laboratory of Photophysics Research and Application, Hebei Normal University, Shijiazhuang, Hebei 050024, China\\
$^2$ Institute of Physics, Chinese Academy of Sciences, Beijing 100190, China\\
$^3$ School of Physical Sciences, University of Chinese Academy of Sciences, Beijing 100049, China\\
$^4$ School of Science Tianjin University of Technology, Tianjin 300384, China\\
$^5$ Beijing Academy of Quantum Information Sciences, Beijing 100193, China}

\begin{abstract}
We present the generalized state-dependent entropic uncertainty relations in multiple measurements setting, and the optimal lower bound is obtained by considering different measurement sequences. We then apply this uncertainty relation to witness entanglement, and give the experimentally accessible lower bounds on both bipartite and tripartite entanglements.
This method of detecting entanglement is applied to a physical system of two particles on a one-dimensional lattice, and GHZ-Werner state.
It is  shown that, for measurements that are not in mutually unbiased bases, this new entropic uncertainty relation is superior to the previous  state-independent one in entanglement detection. Furthermore, we conduct an online experiment to detect multipartite entanglement of GHZ states up to 10 qubits on Quafu cloud quantum computation platform. Our results might play important roles in detecting multipartite entanglement  experimentally.
\end{abstract}

\maketitle

\section{Introduction}
The uncertainty principle sets limits on the precise prediction of the outcomes of two incompatible measurements, which is originally formulated by variance \cite{heisenberg,robertson}.
It is known that entropy is also an uncertainty quantifier, and the entropic uncertainty relation has been constructed in the past decades \cite{uncertainty1,uncertainty2,uncertainty3}.
However, Berta et al. found that the uncertainty can be decreased if the measured system is entangled with a quantum memory, and then they proposed a quantum-memory-assisted entropic uncertainty relation  \cite{berta10nature}.
After that, significant progresses have been made to generalize this entropic uncertainty relation to multiple measurements \cite{hfan15pra,csyu15sr,smfei16pra,smfei18pra,wehner10njp,17rmp,jlchen21pra,Dolatkhah19,Haddadi21,Dolatkhah22}.

The lower bounds of these uncertainty relations are known as state-independent, because the complementary factors come from the overlaps between the eigenstates of measurement operators, which are only connected with measurements.
The characteristic of this kind of uncertainty relation is that they are tight for measurements with mutually unbiased bases (MUBs) \cite{ivonovic1981,s2002}. If the bases of selected measurements are non-MUBs, the inequality is far from tightness.  Recently, B. Bergh and M. G{\"a}rttner proposed a fully-state-dependent entropic uncertainty relation, which gives a tighter lower bound for some measurements with non-MUBs \cite{bergh21prl,bergh21pra}. However, the generalized state-dependent entropic uncertainty relation for multiple measurements needs to be explored.  It remains interesting whether the generalized   uncertainty relation for multiple  measurements with non-MUBs is still tighter than the previous one.

The quantum-memory-assisted entropic uncertainty relation has many applications in quantum information, one of which is to witness entanglement  \cite{berta10nature,bergh21prl,bergh21pra,li11na,prevedel11na,berta14pra,Schneeloch20PRR,xaheng21physik}.
Detection of entanglement in experiment is important in quantum information processing tasks \cite{lukas16prl,zwerger19prl,lu18prx}. However, estimation of entanglement measures requires complete knowledge of quantum state, which is hard to implement in experiment, especially for high-dimensional states and multipartite states \cite{bernd14prl}. Thus, lower bounds are usually presented to evaluate entanglement \cite{dev05,horodecki99pra,chen12prl,bai14prl,nicacio14prl,dai20pra}. The quantum-memory-assisted entropic uncertainty relation provides an experimentally accessible lower bound on bipartite entanglement \cite{bergh21prl,bergh21pra}. It only needs several measurements and does not rely on the  tomography of quantum state. Therefore, one may ask whether the similar method of  obtaining lower bounds on multipartite entanglement can be realized by using the entropic uncertainty relation.

Over the past few years, cloud quantum computation has been available online for various applications, such as tests of fundamental physics  \cite{ychen19pra,wang19npj,mooney19sr,ku20npj,wei20pra,yang22pra,hfan22cp}.
Recently, a newly implemented superconducting cloud quantum computation platform has been launched online, which is named as  ScQ  (available at http://q.iphy.ac.cn),
and then  updated at Quafu cloud quantum computation platform (available at http://quafu.baqis.ac.cn). They have successfully generated a 10-qubit GHZ state and verified its fidelity \cite{hfan22cp}.  It is convenient to carry out the experiment of  multipartite entanglement detection on this cloud platform.

In this paper, we will prove a generalized state-dependent entropic uncertainty relation for multiple measurements with a pair of referenced measurements. We find that the lower bound of the uncertainty relation depends  on the order of measurements, which can be made tighter by taking over all  measurement orders.
As an application, we apply it to multipartite entanglement detection. On the one hand, this new uncertainty relation provides a lower bound on bipartite entanglement, which is shown in a physical model of two particles governed by the Hubbard Hamiltonian.   On the other hand, we give a lower bound on tripartite entanglement  by using this new uncertainty relation, and the  example of GHZ-Werner state is investigated in detail. Finally, we detect multipartite entanglement of GHZ states from 3 to 10 qubits through the Quafu cloud quantum computation platform.

\section{Generalized state-dependent entropic uncertainty relations for multiple measurements}
The quantum-memory-assisted entropic uncertainty relation was first proposed by Berta et al. \cite{berta10nature}, which can be written as
\begin{equation}\label{1}
 S(Q|B)+S(R|B)\geq \log_2 \frac{1}{c} +S(A|B),
\end{equation}
where  $c=\max_{ij}|\langle q_i|r_j\rangle |^2$ with $\{\ket{q_i}\}$ and $\{\ket{r_j}\}$ being eigenvectors of $Q$ and $R$,  $S(A|B)$ is conditional von Neumann entropy of state $\rho_{AB}$, $S(Q|B)$ is the conditional entropy of the post-measurement state
$\rho_{QB}=\sum_i (\ket{q_i}\bra{q_i}\otimes I)\rho_{AB}(\ket{q_i}\bra{q_i}\otimes I)$, and likewise for $S(R|B)$.
If there exist $N$ projective measurements $\{M_m\}_{m=1}^N$ applied on A, the entropic uncertainty relation for multiple measurements has been extended to be \cite{hfan15pra}
\begin{equation}\label{m}
    \sum_{m=1}^N S(M_m|B)\geq -\log_2(b) +(N-1)S(A|B),
\end{equation}
where
\begin{equation}\label{b}
    b=\max_{i_N}
    \left\{\sum_{i_2 \cdots i_{N-1}}\max_{i_1}(|\langle u^1_{i_1}|u^2_{i_2}\rangle|^2) \prod_{m=2}^{N-1}|\langle u^m_{i_m}|u^{m+1}_{i_{m+1}}\rangle|^2\right\},
\end{equation}
with $\ket{u^m_{i_m}}$ being the $i_m$-th eigenstate of $M_m$.

Recently, B. Bergh and M. G{\"a}rttner proposed a fully-state-dependent entropic uncertainty relation  \cite{bergh21prl,bergh21pra},  but the generalization for multiple measurements needs to be studied.  In the following, we will derive the generalized state-dependent entropic uncertainty relation for multiple measurements, which avoids maximization of the measurement bases' overlaps. Now we consider a set of measurements $\{M_i\}_{i=1}^N$ made on subsystem $A$, and label $\ket{\mathbb{M}^i_{m_i}}$ as the $m_i$-th measurement base of $M_i$. Denoting measurement $X$ on $A$ and $Y$ on $B$ as a pair of referenced measurements with respective orthonormal bases $\{\ket{\mathbb{X}_x}\}$ and $\{\ket{\mathbb{Y}_y}\}$, the post-measurement state can be written as
\begin{equation}\label{rhoxy}
\rho_{XY}=\sum_{xy}\Pi_{xy}\rho_{AB}\Pi_{xy},
\end{equation}
with $\Pi_{xy}=\proj{\mathbb{X}_x}\otimes\proj{\mathbb{Y}_y}$.
Before the proof of uncertainty relation, we provide a lemma first.

\textit{Lemma 1}. Given a pair of referenced measurements (i.e. $X$ on $A$ and $Y$ on $B$), let $M_1, M_2, \ldots , M_N$ be $N$ projective measurements made on $A$,  and the following relation holds
\begin{eqnarray}\label{lemma1}
  &&\sum_{i=1}^NS(M_i| B)+H(X|Y)-NS(A|B)-S(\rho_{AB})\nonumber\\
  &&\geq S(\rho_{AB}\|\sum_{m_{\!_N} y}\beta_{m_{\!_N} y}^N\Pi_{m_{\!_N}y}),
\end{eqnarray}
where $\beta_{m_{\!_N} y}^N=\sum_{ xm_1\ldots m_{\!_{N-1}}} c_{xm_1}c_{m_1m_2}\ldots c_{m_{\!_{N-1}}m_{\!_N}}p_{x|y}$ with $c_{xm_1}=|\langle \mathbb{X}_x|\mathbb{M}^1_{m_1}\rangle |^2$ and $c_{m_im_{i+1}}=|\langle \mathbb{M}^{i}_{m_i}|\mathbb{M}^{i+1}_{m_{i+1}}\rangle |^2$, and $\Pi_{m_{\!_N}y}=\proj{\mathbb{M}^N_{m_{\!_N}}}\otimes\proj{\mathbb{Y}_y}$.

This lemma can be proved via an iterative approach \cite{hfan15pra,coles12prl}, by applying  a set of operations
$\{\Lambda_{i}(\rho)=\sum_{m_i} (\ket{\mathbb{M}^i_{m_i}}\bra{\mathbb{M}^i_{m_i}}\otimes I)\rho(\ket{\mathbb{M}^i_{m_i}}\bra{\mathbb{M}^i_{m_i}}\otimes I)\}$  on $\rho_{XY}$, and the proof has been shown in Appendix A.

\textit{Theorem 1}.
 Given a pair of referenced measurements (i.e. $X$ on $A$ and $Y$ on $B$), let $M_1, M_2, \ldots , M_N$ be $N$ projective measurements  made on $A$, and the state-dependent entropic uncertainty relation for multiple measurements reads
\begin{equation}\label{theorem1}
 \sum_{i=1}^N S(M_i|B)+H(X|Y)\geq NS(A|B)+q_1,
\end{equation}
where  $q_1=-\sum_{m_{\!_N}y}p_{m_{\!_N}y}\log_2 \beta_{m_{\!_N}y}^N $ which is called the complementary factor, the parameter $\beta_{m_{\!_N} y}^N$ is the same as lemma 1, and $p_{m_{\!_N}y}=\bra{\mathbb{M}^N_{m_{\!_N}}}\bra{\mathbb{Y}_y}\rho_{AB}\ket{\mathbb{M}^N_{m_{\!_N}}}\ket{\mathbb{Y}_y}$.

\textit{Proof}.---
The relation (\ref{theorem1}) stems from lemma 1 due to the following relation
\begin{eqnarray}\label{S}
  &&S(\rho_{AB}\|\sum_{m_{\!_N} y}\beta_{m_{\!_N} y}^N\Pi_{m_{\!_N} y})   \nonumber\\
  &&= \tr(\rho_{AB}\log_2 \rho_{AB})-\tr(\rho_{AB}\log_2 \sum_{m_{\!_N} y}\beta_{m_{\!_N} y}^N\Pi_{m_{\!_N} y}) \nonumber\\
   &&= -S(\rho_{AB})  -\sum_{m_{\!_N} y}p_{m_{\!_N} y}\log_2 \beta_{m_{\!_N}y}^N.
\end{eqnarray}
We complete the proof of the theorem by substituting Eq. (\ref{S}) into Eq. (\ref{lemma1}). {\qed}

The uncertainty relation (\ref{theorem1}) is derived from iteration based on the conditional entropy of $\rho_{XY}$.
If we consider the $N$ projective measurements first, and apply the operation $\Lambda_{xy}(\rho)=\sum_{xy}  \Pi_{xy}\rho \Pi_{xy}$ at last, then the complementary factor will be
modified in another way.

\textit{Corollary 1}.
Let $M_1, M_2, \ldots , M_N$ be $N$ projective measurements  on $A$.  Let $X$ be the  referenced measurement on $A$ and $Y$ the referenced  measurement on $B$, then
\begin{equation}\label{corollary1}
 \sum_{i=1}^N S(M_i|B)+H(X|Y)\geq NS(A|B)+q_2,
\end{equation}
where $q_2=-\sum_{xy}p_{xy}\log_2 \gamma_{xy}^N$ with the parameter $\gamma_{xy}^N=\sum_{m_1\ldots m_{\!_N}}(\prod_{i=1}^{N-1}c_{m_im_{i+1}})c_{xm_{\!_N}} p_{m_1|y}$.

\textit{Proof}.---
By iteration with $\Lambda_{i}$ at first, an inequality similar to lemma 1 has been obtained in Ref. \cite{hfan15pra}
\begin{equation}\label{q2}
    -NS(A|B)+\sum_{i=1}^NS(M_i| B)    \geq S(\rho_{AB}\parallel\sigma),
\end{equation}
where $\sigma=\sum_{m_1\ldots m_{\!_N}}\prod_{i=1}^{N-1}c_{m_im_{i+1}}\ket{\mathbb{M}^N_{m_{\!_N}}}\bra{\mathbb{M}^N_{m_{\!_N}}}\otimes \bra{\mathbb{M}^1_{m_1}}\rho_{AB}\ket{\mathbb{M}^1_{m_1}}$.
Then we apply the operation $\Lambda_{xy}$ on the right-hand side of Eq. (\ref{q2}) and obtain
\begin{eqnarray}
   && S(\rho_{AB}\parallel\sigma) \nonumber\\
   && \geq S(\Lambda_{xy}(\rho_{AB})\parallel\Lambda_{xy}(\sigma)) \nonumber\\
   &&  =S(\rho_{XY}\parallel\sum_{xy}\sum_{m_1\ldots m_{\!_N}}\prod_{i=1}^{N-1}c_{m_im_{i+1}}c_{xm_{\!_N}} p_{m_1y}\Pi_{xy}) \nonumber\\
   &&=-H(\rho_{XY})-\sum_{xy}p_{xy}\log_2 \sum_{m_1\ldots m_{\!_N}}\prod_{i=1}^{N-1}c_{m_im_{i+1}}c_{xm_{\!_N}} p_{m_1y} \nonumber\\
   && =-H(X|Y)-\sum_{xy}p_{xy}\log_2 \gamma_{xy}^N,
\end{eqnarray}
where $\gamma_{xy}^N=\sum_{m_1\ldots m_{\!_N}}\prod_{i=1}^{N-1}c_{m_im_{i+1}}c_{xm_{\!_N}} p_{m_1|y}$ with $p_{m_1|y}=p_{m_1y}/p_y$.
Therefore, we complete the proof. {\qed}

It is shown that different measurement orders lead to different complementary factors, which is similar to the results in Ref. \cite{csyu15sr}.
Thus we can obtain the following result by taking the optimal measurement sequence.

\textit{Theorem 2}.
Given the referenced measurement $X$ with a set of measurements $\{M_1, M_2, \ldots , M_N\}$ made on $A$ and a referenced measurement $Y$ made on $B$,  let's rearrange the $N+1$ measurements on $A$ in an order $\varepsilon$.
Denoting $\mathcal{E}_i$ as the $i$th measurement in the $\varepsilon$ order with $\{\ket{\mathbb{E}^i_{\varepsilon_i}}\bra{\mathbb{E}^i_{\varepsilon_i}}\}$ being its projectors, the optimal entropic uncertainty
relation gives
\begin{equation}\label{theorem2}
  \sum_{i=1}^N S(M_i|B)+H(X|Y)\geq NS(A|B)+\max_{\varepsilon}q_{\varepsilon},
\end{equation}
where $q_{\varepsilon}=-\sum_{\varepsilon_{\!_{N+1}}y}p_{\varepsilon_{\!_{N+1}}y} \log_2\sum_{\varepsilon_1 \ldots \varepsilon_{\!_{N}}}p_{\varepsilon_1|y}\prod_{i=1}^{N} c_{\varepsilon_i\varepsilon_{i+1}} $ is the complementary factor in measurement order $\varepsilon$, and the maximization runs over all measurement orders $\varepsilon$.

\textit{Proof}.---
Given the order of $N+1$ measurements $\{X, M_1, M_2, \ldots , M_N\}$ made on $A$, it is easy to obtain an uncertainty relation similar to  Theorem 1.
Changing the order of measurements, one can obtain $(N+1)!$ uncertainty relations with different complementary factors. The optimal lower bound is obtained by taking over all measurement orders $\varepsilon$.
{\qed}


\section{Multipartite entanglement detection}
\subsection{Detecting bipartite entanglement}
The entanglement of formation is a useful measure to quantify bipartite entanglement \cite{wootters98prl,horodecki09rmp}, which is defined as
\begin{equation}
    E_f(\rho_{AB})=\min_{\{p_i,\ket{\psi_i}\}} \sum_i p_i S(\tr_B[|\psi_i\rangle\langle\psi_i|]),
\end{equation}
where the minimum is taken over all ensembles $\{p_i, |\psi_i\rangle\}$ satisfying $\rho_{AB}=\sum_i p_i |\psi_i\rangle\langle\psi_i|$.
It's difficult to calculate for general mixed states. However,
it has been known that coherent information $-S(A|B)$ is the lower bound on entanglement of formation \cite{dev05}, which can be used to witness entanglement.
From Eq. (\ref{theorem2}) in Theorem 2, we obtain that
\begin{equation}\label{2}
  -S(A|B)\geq \frac{1}{N}\left[q_m-\sum_{i=1}^N S(M_i|B)-H(X|Y)\right].
\end{equation}
Here the parameter $q_m$ refers to the maximal complementary factor by taking over all measurement sequences. To make the terms on the right-hand side measurable  experimentally, one needs to apply the data-processing inequality, i.e., $H(M_i|M_i^{\prime})\geq S(M_i|B)$, in which
$\{M_i^{\prime}\}_{i=1}^N$ is a set of projective measurements made on $B$.
The lower bound can be obtained experimentally by estimating the complementary factor $q_m$ and  classical conditional entropy (namely, $H(X|Y)$ and  $H(M_i|M_i^{\prime})$), which only needs the probability distribution of measurements.
Thus, the entropic uncertainty relation provides a lower bound
on coherent information, which can be used to detect entanglement directly.

Next we consider an example of two distinguishable particles on a one-dimensional lattice, to estimate the entanglement between two particles.
The Hamiltonian is written as
\begin{equation}\label{H}
    H=-J \sum_{p\in\{A,B\}}\sum_{i=1}^{L-1}(\hat{a}^{\dagger}_{p,i}\hat{a}_{p,i+1}+H.c.)+U\sum_{i=1}^L \hat{n}_{A,i}\hat{n}_{B,i},
\end{equation}
where $L$ lattice sites, $J$ hopping strength, $U$ interaction strength, $\hat{a}^{\dagger}_{p,i} (\hat{a}_{p,i})$  the creation (annihilation) operator for particle $p$ and lattice site $i$,  and $\hat{n}_{p,i}=\hat{a}^{\dagger}_{p,i}\hat{a}_{p,i}$. The Hilbert space of one particle is spanned by ``site basis" $\{\ket{i}\}_{i=1}^L$, which constitute the bases of a natural measurement in this system. The second measurement can be picked as the ``tilted basis" measurement, which is realized by letting the system evolve for a period of time and then measuring the particle's position. The evolution is unitary, which is described  as $R(t)=\exp[itH(J=1,U=0)]$.  We choose the third measurement to be ``tilted basis" measurement with fixed evolution time $t_f=0.38L$. The preparation of the ground state and the realization of the above measurements have been demonstrated in experiment \cite{19nat,16science,20nat}.

\begin{figure}[htbp]
	\begin{center}
		\epsfig{figure=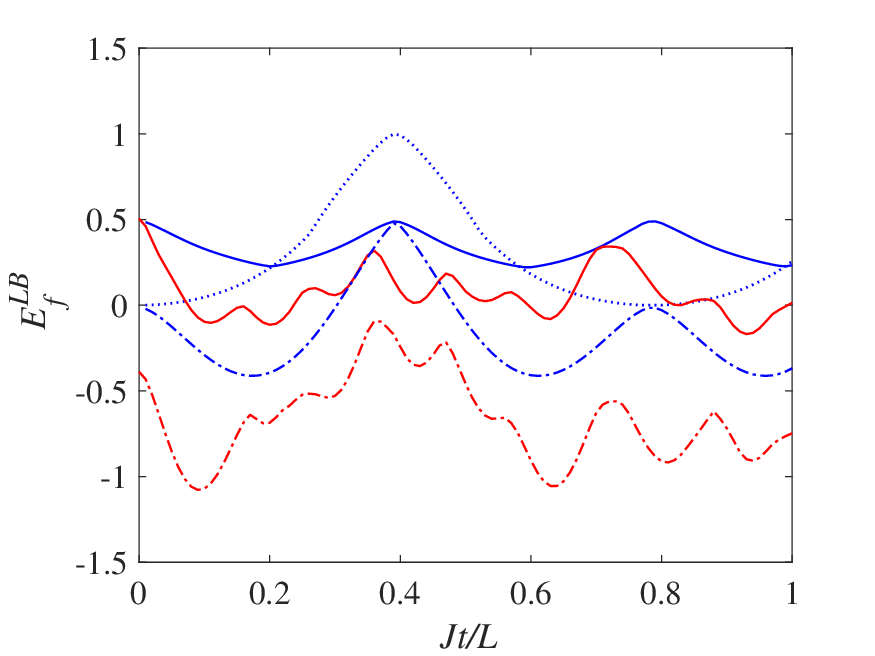,width=0.45\textwidth}
	\end{center}
	\caption{(Color online) The lower bound on entanglement of formation $E_f^{LB}$ of the ground state for the system with $J=1$ and $U=-100$. The blue and red lines correspond to $L=2$ and $L=4$ respectively. Detectable entanglement are plotted by using Eq. (\ref{theorem2}) (solid lines) and Eq. (\ref{m}) (dot-dashed lines) with three measurements, while the blue dotted line is  the case of $L=2$  by using Eq. (\ref{theorem2}) with two measurements.}
\end{figure}

We consider the bipartite entanglement between particles $A$ and $B$ in the ground state of the system with $J=1$ and $U=-100$. The lower bound of entanglement of formation can be evaluated by coherent information, which can be detected by  Eq. (\ref{2}). As shown in Fig. 1, the blue and red solid lines correspond to cases of $L=2$ and $L=4$ respectively, by using entropic uncertainty relation (\ref{theorem2}) with three measurements involving site basis $\{\ket{i}\}$, tilted basis with evolution $R(t)$ and $R(t_f)$. Compared with the results that use Eq. (\ref{m})  to estimate entanglement (dot-dashed lines), our generalized entropic uncertainty relation is more efficient in detecting entanglement.  The blue dotted lines are plotted as the detectable entanglement by means of Eq. (\ref{theorem2}) with the first two measurements in the case of $L=2$, which is consistent with the results in Ref. \cite{bergh21pra,bergh21prl}.
Although the increasing number of measurements may reduce the efficiency of entanglement detection, this new uncertainty relation does better for some  measurements with non-MUBs. For example, in the case of $L=2$,
focusing on the interval of abscissa  $[0,0.2]$ and $[0.6,1]$, at which the  measurements are not in MUBs, the detectable entanglement with three measurements (blue solid line) is higher than that with two measurements (blue dotted line).

\subsection{Detecting tripartite entanglement}
In order to detect tripartite entanglement, we choose the measure of tripartite entanglement of formation that reads   \cite{guo20pra}
\begin{equation}\label{E3}
  E_F^{(3)}(\rho)=\min_{\{p_i,\ket{\psi_i}\}} \sum_i p_i [S(\rho^i_A)+S(\rho^i_B)+S(\rho^i_C)]/3,
\end{equation}
where the minimum runs over all the pure state decompositions $\rho=\sum_i p_i \ket{\psi_i}\bra{\psi_i}$, and  $\rho^i_X$ are the reduced states of subsystem X in state $\ket{\psi_i}$. We choose $1/3$ as the coefficient in Eq.(\ref{E3}) for normalization.
The tripartite entanglement of formation is a multipartite entanglement measure that satisfies complete monogamy relation \cite{guo20pra}.
Consider $\rho_{ABC}=\sum_i q_i \ket{\phi_i}\bra{\phi_i}$ as the optimal pure state decomposition, and  then we have
\begin{eqnarray}
    E_F^{(3)}(\rho)&=&\frac{1}{3} (\sum_i q_i S(\rho^i_A)+\sum_i q_i S(\rho^i_B)+\sum_i q_i S(\rho^i_C)) \nonumber\\
   &=&  \frac{1}{3} (-\sum_i q_i S(A_i|B_iC_i)-\sum_i q_i S(B_i|A_iC_i)\nonumber\\
  && -\sum_i q_i S(C_i|A_iB_i)) \nonumber\\
   &\geq&\frac{1}{3} (- S(A|BC)-S(B|AC)- S(C|AB)),
\end{eqnarray}
where the second equality uses the relation $S(A_i|B_iC_i)=-S(\rho^i_A)$ for pure state $\ket{\phi_i}$, and the last step is due to the fact that conditional entropy is concave.
The above inequality combined with Eq.(\ref{2}) gives the method to detect tripartite entanglement.

\textit{Observation}.
For a tripartite quantum state $\rho_{ABC}$, choosing a sets of measurements $\{M^A_i\}$  with referenced measurement $X$ made on subsystem $A$, and $\{M^B_i\}$ with $Y$ on $B$, and $\{M^C_i\}$ with $Z$ on $C$, the tripartite entanglement of formation can be detected by
\begin{eqnarray}\label{observation}
  E_F^{(3)}&\geq& \frac{1}{3N}[q_m-\sum_{i=1}^N H(M^A_i|M^B_iM^C_i)-H(X|YZ)  \nonumber\\
  &+&q_m^{\prime}- \sum_{i=1}^N H(M^B_i|M^A_iM^C_i)-H(Y|XZ)  \nonumber\\
  &+&q_m^{\prime\prime} - \sum_{i=1}^N H(M^C_i|M^A_iM^B_i)-H(Z|XY)]  ,
\end{eqnarray}
 where  $H(\cdot|\cdot)$ is the classical conditional entropy, and $q_m, q_m^{\prime}, q_m^{\prime\prime}$ are the optimal complementary factors of the corresponding uncertainty relations.

\begin{figure}[htbp]
	\centering
	\epsfig{figure=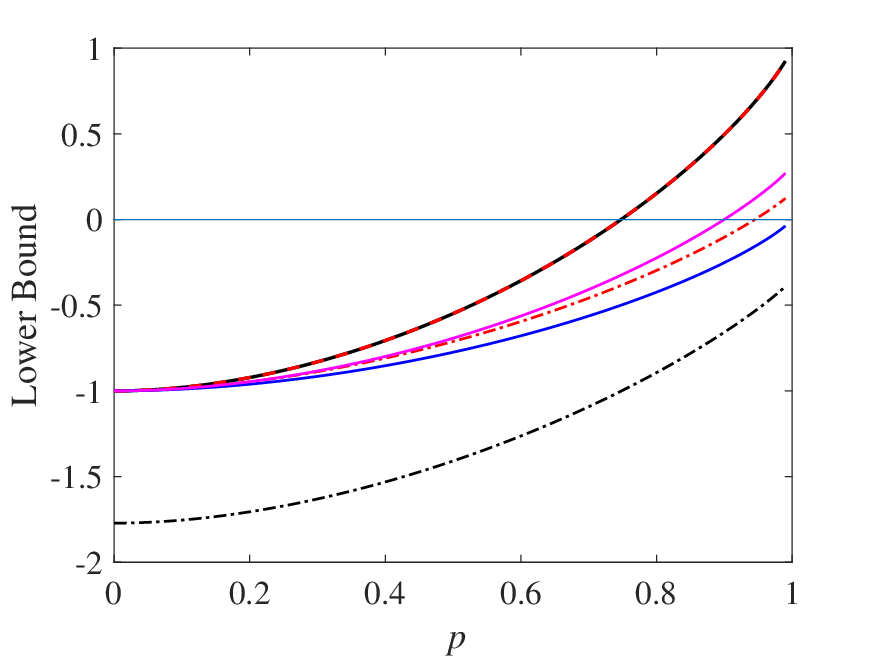,width=0.45\textwidth}
	
	\caption{(Color online) The lower bound of $E_F^{(3)}$ for GHZ-Werner state. The black and red dashed lines give the lower bound by using the uncertainty relations (\ref{m}) and (\ref{theorem2}) respectively with the same set of measurements $\{\sigma_z,\sigma_x\}$. The black and red dot-dashed lines are plotted by using the uncertainty relations (\ref{m}) and (\ref{theorem2}) respectively with two measurements $\{\sigma_z,\sigma_r\}$. The blue and purple curves are obtained by Eq. (\ref{theorem2}) with respective set of measurements $\{\sigma_x,\sigma_y,\sigma_z\}$ and $\{\sigma_x,\sigma_z,\sigma_r\}$.}
\end{figure}

We   consider the GHZ-Werner state to estimate the lower bound of $E_F^{(3)}$ via relation (\ref{observation}). The three-qubit state is defined as
\begin{equation}\label{GHZ}
    \rho_{p}=p\ket{{\rm GHZ}}\bra{{\rm GHZ}}+(1-p)\frac{\mathbf{I}}{8},
\end{equation}
where $\ket{{\rm GHZ}}=(\ket{000}+\ket{111})/\sqrt{2}$ is the GHZ state, $\mathbf{I}$ is the identity matrix, and $0\leq p\leq 1$.  Choosing $\sigma_x$ and $\sigma_z$ as two measurements made on each qubit, the lower bounds of tripartite entanglement using both Eqs. (\ref{m}) and (\ref{theorem2})  are shown in Fig. 2. The results are the same (i.e. the black and red dashed lines coincide), which indicates that when $p>0.747614$, tripartite entanglement can be detected in GHZ-Werner state. However, if we choose measurements $\sigma_z$ and $\sigma_r=(\sigma_z+\sigma_x)/\sqrt{2}$, which are not in mutually unbaised bases, the entanglement detection efficiency is reduced. In spite of this, the lower bound calculated by means of our relation (\ref{theorem2}) (red dot-dashed line) is better than the results with the previous uncertainty relation (\ref{m}) (black dot-dashed line). That is to say, for non-MUBs measurements, the state-dependent uncertainty relation (\ref{theorem2}) is more effective to detect tripartite entanglement due to the varying overlaps, which is consistent with the results of bipartite entanglement detection \cite{bergh21prl}.  This conclusion holds true for multiple measurements. As shown in Fig. 2, the blue and purple curves correspond to the cases  with respective set of measurements $\{\sigma_x,\sigma_y,\sigma_z\}$ and $\{\sigma_x,\sigma_z,\sigma_r\}$ by using Eq. (\ref{theorem2}). The latter is better than the former, and even better than the case with two measurements $\{\sigma_z,\sigma_r\}$ (red dot-dashed line).

Note that a scheme to quantify genuine tripartite entanglement with entropic correlations has been recently proposed  \cite{Schneeloch20PRR}, where
the genuine multipartite entanglement measure (GMEM) based on entanglement of formation can be estimated by the following inequality \cite{Schneeloch20PRR,Szalay15pra}
\begin{equation}\label{V}
  E^{GME}_{f}\geq -S(A|BC)-S(B|AC)-S(C|AB)-2\log_2(d_{\max}),
\end{equation}
where $d_{\max}$ is the maximal dimension of subsystems $A$, $B$ and $C$.
Therefore, if the bases of chosen measurements are non-MUBs, the genuine tripartite entanglement can be detected more effectively by applying this new uncertainty relation in a similar way.

\subsection{Generalization in multipartite systems}
The procedure of entanglement detection can be extended
 to a general multipartite  systems $\rho_{A_1\ldots A_m}$. The $m$-partite entanglement measure based on entanglement of formation is defined as \cite{guo20pra} $E_F^{(m)}=\min_{\{p_i,\ket{\psi_i}\}} \sum_i p_i E_F^{(m)}(\ket{\psi_i})$, in which $E_F^{(m)}(\ket{\psi_i})=\sum_k S(\rho^i_{A_k})/m$, with $\rho^i_{A_k}$ being reduced  state of $k$-th subsystem $A_k$ for $\ket{\psi_i}$. The lower bound of general $m$-partite entanglement is derived as follows
\begin{equation}\label{multipartite_entanglement}
    E_F^{(m)}\geq \frac{1}{m} \sum_{k=1}^m -S(\rho_{A_k}|\rho_{\tilde{A}_k}),
\end{equation}
where $\rho_{\tilde{A}_k}$ is the reduced state after tracing out the subsystem $A_k$. Here we also choose $1/m$ as the coefficient for normalization. The right hand side of the above inequality can be estimated by entropic uncertainty relation in a similar way as shown in Eq. (\ref{2}). Therefore, multipartite entanglement can be detected by means of this method as well.

\section{Entanglement detection of GHZ states on  Quafu cloud quantum computation platform}
\begin{figure}[htbp]
	\centering
	\epsfig{figure=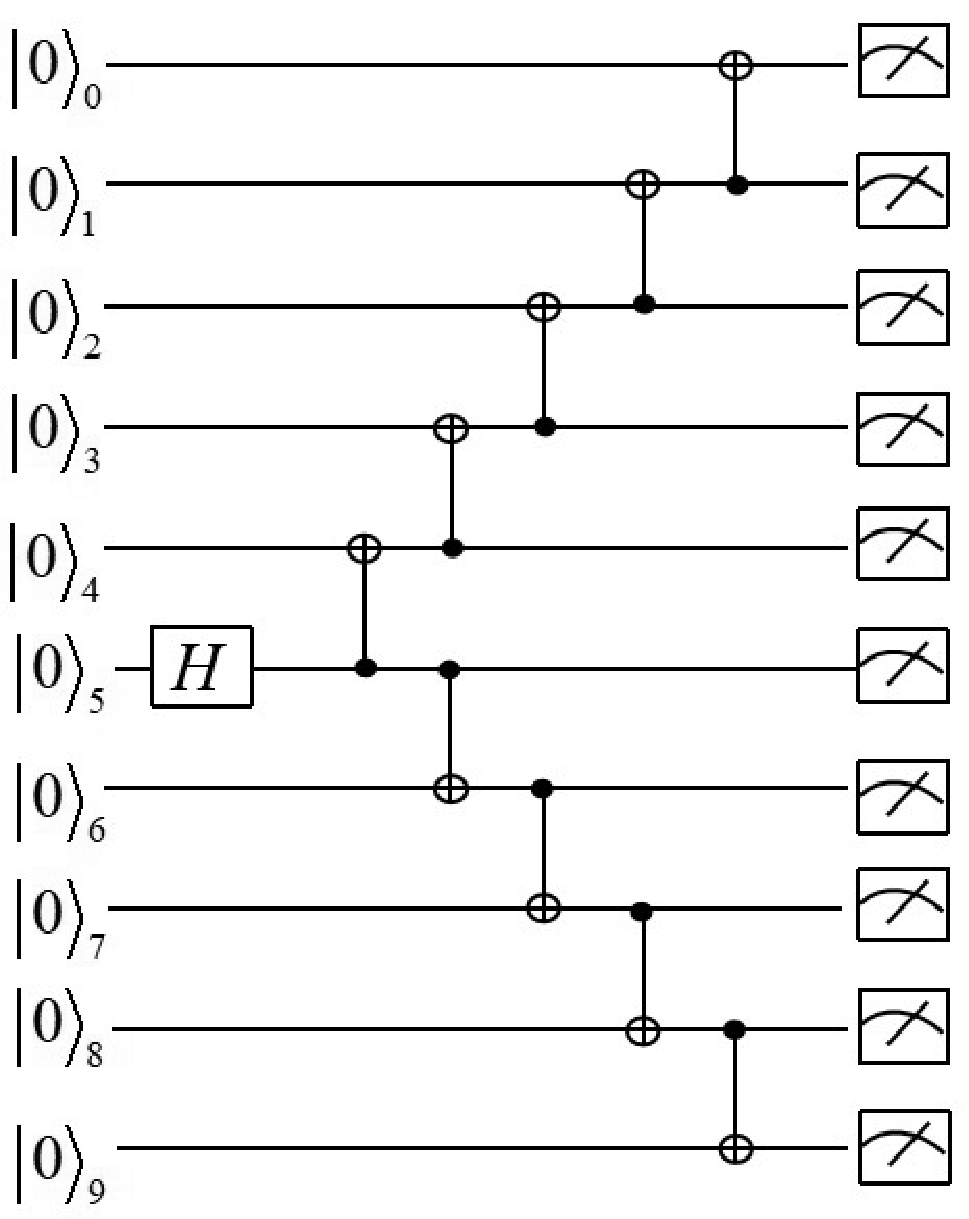,width=160pt, height=200pt}
	
	\caption{(Color online) Quantum circuit of generating 10-qubit GHZ state.}
\end{figure}

\begin{figure*}[htbp]
	\centering
	\epsfig{figure=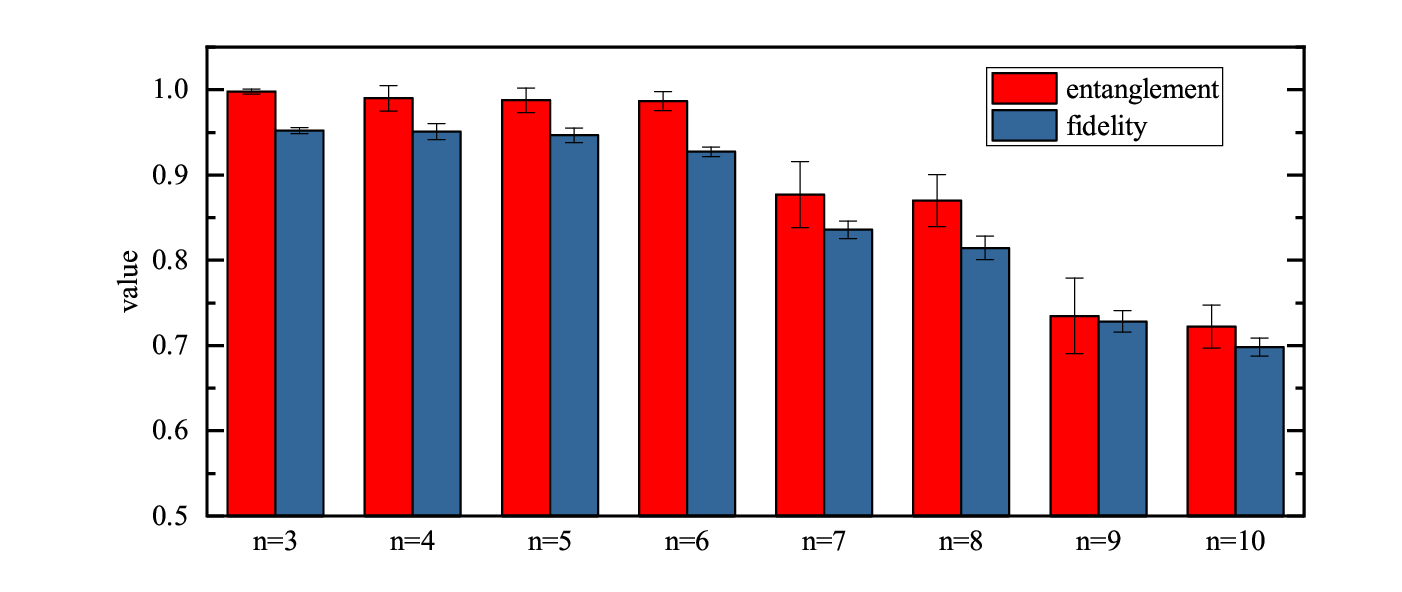,width=0.88\textwidth}	
	\caption{(Color online) The lower bound on $E_F^{(m)}$ and fidelity of  n-qubit GHZ  states. From left to right, the qubits used are $Q_1$-$Q_3$, $Q_1$-$Q_4$, $Q_1$-$Q_5$, $Q_1$-$Q_6$, $Q_1$-$Q_7$, $Q_1$-$Q_8$, $Q_1$-$Q_9$, $Q_1$-$Q_{10}$.}
\end{figure*}

The present backend devices of Quafu cloud quantum computation platform include one 10-qubit processor (ScQ-P10), one 18-qubit processor (ScQ-P20) and one 50-qubit processor (ScQ-P50).
In our experiment, we take the ScQ-P20 as the backend, which allows single-qubit gates with fidelity $>99\%$ and two-qubit CZ gates with fidelity $95\%\sim 99.14\%$. In this section, we will evaluate multipartite entanglement of GHZ states up to 10 qubits.
The circuit of preparing 10-qubit GHZ state is shown in Fig. 3, and the circuits of generating GHZ states from 3 to 9 qubits are designed in a similar way, in which the Hadamard gate acts on the $(n//2)$-th qubit, followed by a series of CNOT gates.

To detect multipartite entanglement $E_F^{(m)}$, we perform two kinds of measurements (namely, $\sigma_z$ and $\sigma_x$), which is the optimal strategy to detect entanglement of GHZ states via the method of entropic uncertainty relations. The results are obtained by averaging
5000 repeated single-shot measurements.  The lower bound of $E_F^{(m)}$  can be obtained by Eq. (\ref{multipartite_entanglement}) combined with Eq. (\ref{2}), which only needs the  probability distribution  of two measurements. The detectable entanglement of n-qubit GHZ states described by the lower bound of $E_F^{(m)}$  are calculated as $0.9980\pm 0.0030$ (n=3), $0.9899\pm 0.0148$ (n=4), $0.9875\pm 0.0144$ (n=5), $0.9868\pm 0.0112$ (n=6), $0.8771\pm 0.0386$ (n=7), $0.8699\pm 0.0305$ (n=8), $0.7347\pm 0.0443$ (n=9) and $0.7222\pm 0.0252$ (n=10), which have been illustrated in Fig. 4. The error bar corresponds to the standard deviation obtained from eight experiments.
Furthermore, when n=3, the genuine tripartite entanglement can be evaluated by Eq. (\ref{V}). The lower bound of $E^{GME}_{f}$ reaches 0.9940 in 3-qubit GHZ state via this method.

We compare the above results with GHZ fidelity. Here the fidelity of GHZ states are defined as $F=(P+C)/2$, where $P$ comes from the summation of measurement probabilities  $P_{00\cdots 0}$
and $P_{11\cdots 1}$, and $C$ can be obtained by measuring the parity oscillations \cite{hfan22cp,monz11prl,song17prl,song19science,omran19science}.
As shown in Fig. 4, there's not much difference between the general multipartite entanglement detected by  entropic uncertainty relation and GHZ fidelity, but the former is a little higher than the latter.

\section{Discussion and conclusion}
In conclusion, we propose a generalized  entropic uncertainty relation for more than two measurements which is state-dependent due to a pair of referenced measurements made on both system and quantum memory. In multiple measurements setting, the lower bound of uncertainty relation rely on the sequence of measurement, and we give the optimal lower bound by taking over all measurement sequences. We highlight that the quantum-memory-assisted entropic uncertainty relation can be used to detect multipartite entanglement, which requires only the probability distribution of several measurements. By illustrating the examples of ground state of Hubbard model and  GHZ-Werner state, we show that for measurements with non-MUBs, this new uncertainty relation can detect entanglement more efficiently than the previous state-independent one. If the measurements with MUBs are difficult to implement experimentally due to the characteristics of systems or environmental noise,  our results are of importance to improve the efficiency of entanglement detection.
Finally, we use this method to verify multipartite entanglement of  GHZ states up to 10 qubits on Quafu cloud quantum computation platform, and the detectable entanglement evaluated by means of entropic uncertainty relation only needs the probability distribution of measurements $\sigma_z$ and $\sigma_x$, which can be easily acquired in experiment.

\section{Acknowledgments}
This work was supported by the NSF-China (Grants Nos. 12105074 and 12004280),  Hebei NSF (Grants Nos.  A2019205263), and the fund of
Hebei Normal University (Grants Nos. L2019B07).

\appendix
\section{Proof of lemma 1}
\textit{Lemma 1}. Given a pair of referenced measurements (i.e. $X$ on $A$ and $Y$ on $B$), let $M_1, M_2, \ldots , M_N$ be $N$ projective measurements  on $A$,  and the following relation holds
\begin{eqnarray}
  &&\sum_{i=1}^NS(M_i| B)+H(X|Y)-NS(A|B)-S(\rho_{AB})\nonumber\\
  &&\geq S(\rho_{AB}\|\sum_{m_{\!_N} y}\beta_{m_{\!_N} y}^N\Pi_{m_{\!_N}y}),
\end{eqnarray}
where $\beta_{m_{\!_N} y}^N=\sum_{ xm_1\ldots m_{\!_{N-1}}} c_{xm_1}c_{m_1m_2}\ldots c_{m_{\!_{N-1}}m_{\!_N}}p_{x|y}$ with $c_{xm_1}=|\langle \mathbb{X}_x|\mathbb{M}^1_{m_1}\rangle |^2$ and $c_{m_im_{i+1}}=|\langle \mathbb{M}^{i}_{m_i}|\mathbb{M}^{i+1}_{m_{i+1}}\rangle |^2$, and $\Pi_{m_{\!_N}y}=\proj{\mathbb{M}^N_{m_{\!_N}}}\otimes\proj{\mathbb{Y}_y}$.

\textit{Proof}.---
This can be proved using an iterative approach \cite{hfan15pra,coles12prl}. First, we prove the relation for $N=1$. Let $M_1$ be the measurement on system A with projectors labeled by $\{\proj{\mathbb{M}^1_{m_1}}\}$.  Then
\begin{eqnarray}\label{ss}
  &&H(X|Y)-S(A|B) \nonumber\\
  &&= H(\rho_{XY})-H(Y)-S(\rho_{AB})+S(\rho_B) \nonumber\\
   &&= S(\rho_{AB}\parallel \rho_{XY})-H(Y)+S(\rho_B) \nonumber\\
   && \geq  S(\Lambda_{1}(\rho_{AB})\parallel\Lambda_{1}( \rho_{XY}))-H(Y)+S(\rho_B) \nonumber\\
   && = S(\rho_{M_1B}\parallel\sum_{xym_1}c_{xm_1}p_{xy}\Pi_{m_1y}) -H(Y)+S(\rho_B)\nonumber\\
   && = -S(M_1|B)-\sum_{m_1y}p_{m_1y}\log_2 (\sum_{x}c_{xm_1}p_{xy})-H(Y),\nonumber\\
   &&
\end{eqnarray}
where we denote $H(\cdot|\cdot)$ as the classical conditional entropy, the definition and contractive property of relative entropy are used in the the second the third steps,  the measurement operation $\Lambda_{1}(\rho)=\sum_{m_1} (\ket{\mathbb{M}^1_{m_1}}\bra{\mathbb{M}^1_{m_1}}\otimes I)\rho(\ket{\mathbb{M}^1_{m_1}}\bra{\mathbb{M}^1_{m_1}}\otimes I)$ are used in the fourth step with $c_{xm_1}=|\langle \mathbb{X}_x|\mathbb{M}^1_{m_1}\rangle |^2$ and $p_{xy}=\bra{\mathbb{X}_x}\bra{\mathbb{Y}_y}\rho_{AB}\ket{\mathbb{X}_x}\ket{\mathbb{Y}_y}$, and in the last step  $p_{m_1y}=\tr(\rho_{M_1B}\Pi_{m_1y})$ with $\Pi_{m_1y}=\proj{\mathbb{M}^1_{m_1}}\otimes\proj{\mathbb{Y}_y}$.
Since $p_{xy}=p_{x|y}p_{y}$, we have $\log_2 (\sum_{x}c_{xm_1}p_{xy})=\log_2 (\sum_{x}c_{xm_1}p_{x|y})+\log_2 p_{y}$. Then
\begin{eqnarray}\label{hs}
   && -\sum_{m_1y}p_{m_1y}\log_2 (\sum_{x}c_{xm_1}p_{xy}) \nonumber\\
   &&  = -\sum_{m_1y}p_{m_1y}\log_2 (\sum_{x}c_{xm_1}p_{x|y})-\sum_{m_1y}p_{m_1y}\log_2 (p_{y})\nonumber\\
   &&  =-\tr( \rho_{AB}\log_2(\sum_{xym_1}c_{xm_1}p_{x|y}\Pi_{m_1y})) \nonumber\\
   &&\quad +H(Y)+S(\rho_{AB})-S(\rho_{AB}) \nonumber\\
   && =S(\rho_{AB}\|\sum_{xym_1}c_{xm_1}p_{x|y}\Pi_{m_1y}) \nonumber\\
   &&\quad +S(\rho_{AB})+H(Y).
\end{eqnarray}
Combining Eq. (\ref{ss}) and Eq. (\ref{hs}) gives
\begin{eqnarray}\label{m1}
  &&S(M_1| B)+H(X|Y)-S(A|B)-S(\rho_{AB})\nonumber\\
  &&\geq S(\rho_{AB}\|\sum_{xym_1}c_{xm_1}p_{x|y}\Pi_{m_1y}).
\end{eqnarray}
By using the contractive property of relative entropy with operation $\Lambda_{2}(\rho)=\sum_{m_2} (\ket{\mathbb{M}^2_{m_2}}\bra{\mathbb{M}^2_{m_2}}\otimes I)\rho(\ket{\mathbb{M}^2_{m_2}}\bra{\mathbb{M}^2_{m_2}}\otimes I)$, then
\begin{eqnarray}
  &&S(\rho_{AB}\|\sum_{xym_1}c_{xm_1}p_{x|y}\Pi_{m_1y}) \nonumber \\
  && \geq  S(\Lambda_{2}(\rho_{AB})\|\Lambda_{2}(\sum_{xym_1}c_{xm_1}p_{x|y}\Pi_{m_1y})) \nonumber \\
   && =- S(\rho_{M_2B})+S(\rho_B)-S(\rho_B)+S(\rho_{AB})-S(\rho_{AB}) \nonumber \\
   && \quad -\tr(\rho_{M_2B}\log_2 \sum_{xym_1m_2}c_{xm_1}c_{m_1m_2}p_{x|y}\Pi_{m_2y}) \nonumber\\
   && =S(\rho_{AB}\|\sum_{xym_1m_2}c_{xm_1}c_{m_1m_2}p_{x|y}\Pi_{m_2y})\nonumber\\
   &&\quad - S(M_2|B)+S(A|B),
\end{eqnarray}
in which $\Pi_{m_2y}=\proj{\mathbb{M}^2_{m_2}}\otimes\proj{\mathbb{Y}_y}$. Thus,
\begin{eqnarray}\label{m2}
  &&\sum_{i=1}^2S(M_i| B)+H(X|Y)-2S(A|B)-S(\rho_{AB})\nonumber\\
  &&\geq S(\rho_{AB}\|\sum_{xym_1m_2}c_{xm_1}c_{m_1m_2}p_{x|y}\Pi_{m_2y}).
\end{eqnarray}
By iteration with $\Lambda_{i}$, we can get
\begin{eqnarray}\label{mn}
  &&\sum_{i=1}^NS(M_i| B)+H(X|Y)-NS(A|B)-S(\rho_{AB})\nonumber\\
  &&\geq S(\rho_{AB}\|\sum_{xy m_1\ldots m_{\!_N}}c_{xm_1}c_{m_1m_2}\ldots c_{m_{\!_{N-1}}m_{\!_N}}p_{x|y}\Pi_{m_{\!_N}y}). \nonumber
\end{eqnarray}
Thus we complete the proof. {\qed}

\end{document}